\documentclass{emulateapj} 
\usepackage{enumerate,multirow}
\usepackage{amsmath,amssymb,graphicx,natbib,subfigure,dcolumn,bm,comment} 
\usepackage{amsmath}
\usepackage{amssymb} 
\usepackage{color}
\usepackage{wasysym}

 \usepackage{ulem}

\shorttitle{Origin of Hot and Warm Jupiters}
\shortauthors{Antonini, Hamers and Lithwick}

\begin{document}
\def\gap{\;\rlap{\lower 2.5pt
\hbox{$\sim$}}\raise 1.5pt\hbox{$>$}\;}
\def\lap{\;\rlap{\lower 2.5pt
 \hbox{$\sim$}}\raise 1.5pt\hbox{$<$}\;}

\def\yl#1{{\textcolor{blue}{[[\bf YL: #1]]}}}
\def\mb#1{{\textcolor{red}{[[\bf MB: #1]]}}}

\title{Dynamical constraints on the origin of hot and warm Jupiters
  with close friends}

\author{Fabio Antonini$^{1}$, Adrian S. Hamers$^{2}$ and Yoram Lithwick$^{1}$}
\affil{(1) Center for Interdisciplinary Exploration and Research in Astrophysics (CIERA)
and Department of Physics and Astrophysics, Northwestern University,
Evanston, IL 60208;
(2) Leiden Observatory, Niels Bohrweg 2, Leiden, 2333CA, The Netherlands}

\begin{abstract}
 Gas giants orbiting their host star within
the ice line are thought to have 
migrated to their current locations
from farther out.
Here we consider  the origin and dynamical evolution of 
observed Jupiters, focusing on hot and warm Jupiters with outer
friends.
We show that the majority of the observed Jupiter pairs (twenty out of twenty-four) 
will be dynamically unstable if the inner planet was placed at  $\gtrsim
1\rm AU$ distance from the stellar host. 
This finding is at odds with formation theories that invoke the
migration of such planets from semi-major axes 
$\gtrsim 1\rm AU$ due to secular dynamical processes (e.g., secular chaos, Lidov-Kozai oscillations) coupled with tidal dissipation. 
In fact, the results of $N-$body integrations
show that the evolution of dynamically unstable
systems does not lead  to tidal migration but rather to planet
ejections and collisions with the host star.
This and other arguments lead us to suggest
that most of the observed planets with a companion could not have been transported from
further out through secular migration processes.
More generally, by using a combination of numerical and analytic
techniques  we show that the high-$e$ Lidov-Kozai migration scenario
 can only account  for less than  $10\%$ of all 
 gas giants observed between $0.1-1\rm AU$. 
Simulations of multi-planet systems support this result.
Our study indicates that rather than starting on highly eccentric
orbits with orbital periods above one year, these ``warm'' Jupiters are more likely to have
reached the region where they are  observed today without having
experienced significant tidal dissipation.
\end{abstract}

\section{Introduction}The observed abundance of Jupiter-size planets orbiting interior to
the ice-line  around their stars poses a challenge to our current
understanding of planet formation \citep{2012ApJ...753..160W}.
Close-in planets   ($\lesssim 1\rm AU$)  are typically
 thought to have formed beyond the ice-line  where large, icy
cores can grow and accrete, and to
have moved within $\rm 1 AU$ later on.
Possible mechanisms for migration invoke
gentle disk migration \citep[e.g.,][]{1980ApJ...241..425G,1996Natur.380..606L} or tidal interaction 
with the stellar host that gradually removes energy from the planet
orbit. In this latter model the interaction with an external perturber (e.g., a star, a planet
companion) moves the planet  onto a highly eccentric
orbit so that efficient tidal circularization can take 
place\  \citep[e.g.,][]{1996Sci...274..954R,2003ApJ...589..605W,2011ApJ...735..109W,2011Natur.473..187N}. 

Migration scenarios must account for the existence of  both  hot Jupiters
(HJs; gas giants orbiting within 0.1 AU of their host stars) and warm 
Jupiters (WJs; orbiting in the region from 0.1 to 1 AU).
WJs are giant planets observed  within the so called  period ``valley'',
corresponding to the dip in the giant planet orbital period 
distribution from roughly  $P=10$ to  $100\rm days$   \citep{2015arXiv151100643S}.
Thus, WJs are  interior to both the ice-line  
and the observed pileup of giant planets beyond $1\rm $AU. 
While most HJs have nearly zero eccentricities, WJs have a range of
eccentricities with a few being on highly  eccentric ($\gtrsim 0.8$)
orbits   \citep[e.g.,][]{2013ApJ...767L..24D}. 

Although HJs and WJs appear to be separated in their period and
eccentricity  distributions it has been suggested that they might share a common origin. 
A possibility is that  both HJs and WJs migrated inward through high-$e$ migration
processes such as secular chaos and Lidov-Kozai~(LK) cycles coupled with
tidal friction
\citep[e.g.,][]{2014ApJ...781L...5D,2014Sci...346..212D,2016MNRAS.455.1538F}. 
In this scenario the HJ orbits have been fully circularized by tidal friction,
while WJs are still on their way to
become HJs and are experiencing large amplitude 
eccentricity  oscillations induced by an external perturber.
In fact, most gas giants observed in the period valley  have observed
eccentricities that are too small for significant tidal evolution,
but this can be understood if they are currently  near  the low-$e$
phase of a LK cycle, while periodically attaining high eccentricities
and thereby experiencing  significant tidal dissipation.
 
In this paper we examine whether high-$e$ migration models are
consistent with the observed properties of the WJ population. 
We based our analysis on (mostly) radial-velocity  data from the exoplanet database at
http://exoplanets.org  \citep{2011PASP..123..412W}. In particular, we focus on
planets that have a detected  outer companion and that orbit
their stellar host interior to $1\rm AU$. We use both a high
precision three-body integrator as well as an orbit average secular code to
produce synthetic populations of migrating planets. By comparing
our results to observations we are able to address whether the giant planets observed in the period valley could
have formed through secular migration processes. 

We find that secular
processes do cause giant planets to migrate 
within the radial range $0.1-1\rm AU$, however the orbital properties of the
migrating planets are not consistent with what is observed. Our results are
consistent with less than  $10\%$ of all  gas giants observed in the period
valley having migrated through tidal dissipation.
We note that our results are somewhat  complementary to those of
 \citet{2016arXiv160105095H}. These  authors recently used Kepler transit data to show that 
HJs and WJs are  distinct in their respective fractions of
sub-Jovian  companions. They found that HJs as a whole 
do not have any detectable inner or outer planetary companions with
periods inward of 50 days. In stark contrast, half of the WJs in
their sample have  small companions. Motivated by this discovery and 
by additional arguments,   
\citet{2016arXiv160105095H} proposed that a large fraction of WJs are formed in-situ.

The paper is organized as follows.
In Section~2 we consider all planets within $1\rm AU$ that have a
detected  outer companion and address the dynamical stability of these systems.
In Section~3 we describe our numerical methods. In Section~4 we describe
the results of the $N$-body integrations that we used to study the
dynamical evolution of systems close to the stability boundary.
In Sections~5 and 6 we study the dynamical evolution of planets undergoing secular
migration and their resulting orbital distribution. 
Section~7 summarizes our main results.

\section{ Stability}\label{s1}
In this Section
we discuss the stability of observed
    Jupiter pairs hosting HJs and WJs. In particular, we examine
whether the HJ or WJ could
    have reached its current orbit via high-e migration, or whether
    its having a high-$e$  and $a\sim 1\rm AU$ in the past would instead have made the system
dynamically unstable.

The majority of WJs are far enough from their stellar hosts
that they are not expected to experience significant tidal dissipation.
However, if the eccentricity of the WJs are experiencing large amplitude 
LK oscillations induced by an external perturber, then
they might be currently at the low-$e$ phase of a LK cycle. Over a secular timescale
they might access a periapsis separation such that  $a_1(1-e_1^2)<a_{\rm cr}\approx 0.1\rm AU$, within which tidal 
dissipation will cause efficient migration. In this scenario
the WJs have to be accompanied  by a strong enough perturber to overcome 
Schwarzschild precession. Note that at $\sim 0.1\rm AU$, the additional precession
due to tides are negligible compared to Schwarzschild precession for typical hosts.

\citet{2014ApJ...781L...5D} consider the secular migration scenario for
WJs. At the quadrupole level of approximation 
they derive an analytic upper limit on the outer perturber separation
by requiring the WJ to reach $a_1(1-e_1^2)<a_{\rm cr}$ during LK oscillations:
\begin{eqnarray}\label{d+14}
  \frac{a_2\sqrt{1-e_2^2}}{a_{\rm 1}}\lesssim
  \left(\frac{8GM_{\star}}{c^2a_{\rm 1}}\right)^{-1/3} 
\left(\frac{M_\star}{M_2} \right)^{-1/3} ~~~~~~~~~~~~
\\ \left[2e_{\rm 1}^2+3\left(1-\frac{a_{\rm cr}}{a_{\rm 1}}\right) \right]^{1/3} \left(\sqrt{\frac{a_{\rm 1}}{a_{\rm cr}}}-\frac{1}{\sqrt{1-e_{\rm 1}^2}} \right)^{-1/3}
\nonumber
\end{eqnarray}
where $M_\star$ is the mass of the host star, $M_2$ the mass of the outer perturber and
$a_{\rm 1}$ ($a_{\rm 2}$) and $e_{\rm 1}$  ($a_{\rm 2}$)  are the semi-major axis and eccentricity
of the inner (outer) planet.
In the limit $a_{\rm 1}\gg a_{\rm cr}$ and $e_{\rm 1}\rightarrow 0$, Equation\ (\ref{d+14})
becomes
\begin{eqnarray}\label{d+14-2}
\frac{a_2\sqrt{1-e_2^2}}{a_{\rm 1}} &\lesssim&
  \left(\frac{8GM_{\star}}{c^2\sqrt{a_{\rm 1}a_{\rm cr}}}\right)^{-1/3} 
\left(\frac{M_\star}{M_2} \right)^{-1/3} \approx 20 \left(\frac{M_2}{M_{\rm Jupiter}}\right)^{1/3} \ \\
&&\times \left(\frac{M_\star}{M_\odot} \right)^{-2/3}\left(\frac{a_{\rm 1}}{0.2 \rm AU}\right)^{1/6}
\left(\frac{a_{\rm cr}}{0.1 \rm AU} \right)^{1/6}\ .
\nonumber
\end{eqnarray} 
Using the above equations \citet{2014ApJ...781L...5D} concluded that ``for a WJ
at $0.2\rm AU$, a Jupiter perturber is required at $\lesssim 3\rm AU$''.

\begin{figure*}
\centering
\includegraphics[width=3.in,angle=270.]{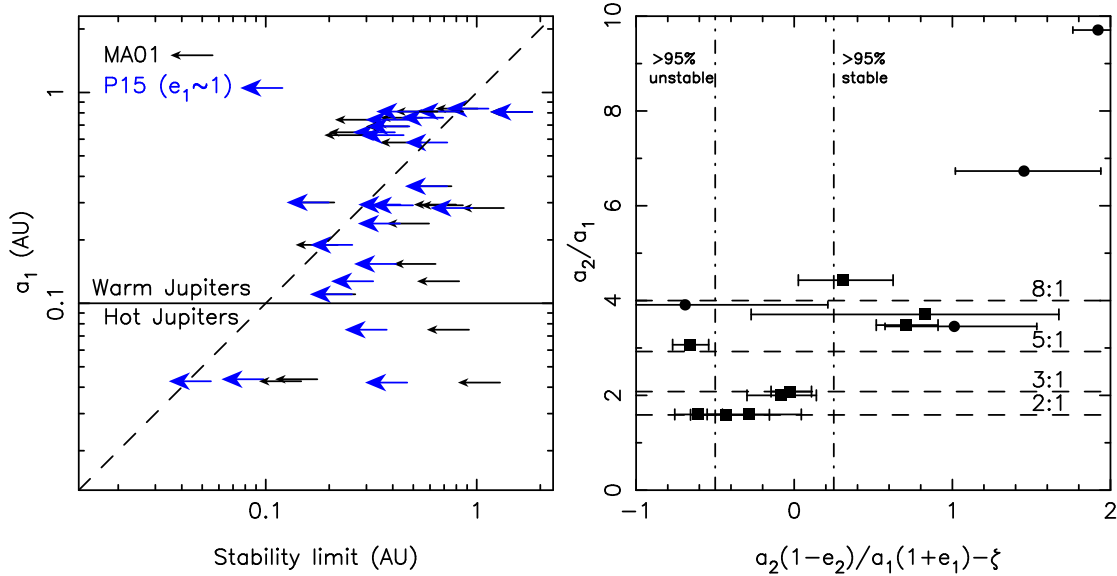}
\caption{The left panel gives the semi-major axis of  HJs and WJs with observed
 companions plotted as a function of the inner planet
 semi-major axis above which the system will be unstable.
 The inner planet must have been to the left of the tip of each arrow during its high e migration
 (assuming it formed via high e migration);
 otherwise,  it would have been dynamically unstable
   according to
 Equation (\ref{MA01}) (black arrows) and Equation\ (\ref{pet+15})
 (blue arrows).  The stability limit imposed by  Equation\ (\ref{pet+15}) was computed taking the
limit $e_1\rightarrow 1$ and adding a factor 0.5 to the right hand
side which approximately corresponds to $95\%$ chance for a system to
be unstable over  $10^8$ years of evolution. 
Systems that are at the left of the dashed line are dynamically
unstable according to the stability criteria we considered.
The right panel shows the stability boundary in
Equation\ (\ref{pet+15}) as a function of the semi-major axis ratio $a_2/a_1$ for
our sample of two-planet systems; here
$\zeta=2.4[\max(\mu_{2},\mu_{1})]^{1/3}\sqrt{a_2 /
  a_1}+1.15$.
The vertical dot-dashed line indicates the region for which 
$> 95\%$ of the systems to the left (right) are expected to be 
unstable (stable). The horizontal dashed lines indicate the position
of some of the strongest mean-motion resonances.
Square (circle) symbols are systems with semi-major axis $a_1\geq
0.5\rm AU$
($\leq 0.5\rm AU$).}  \label{fig1}
\end{figure*}

Previous work  did not consider the stability of the initial configurations that
can lead to the formation of a WJ in the secular migration scenario.
In addition to the condition Equation\ (\ref{d+14}) one must require the planetary system to be 
dynamically stable in its initial configuration, i.e., before tidal dissipation has 
significantly shrank the orbit of the inner planet. 

We compare the observed systems configurations to 
various stability criteria.
We consider the criterion\ \citep{2015ApJ...808..120P}
\begin{equation}\label{pet+15}
\frac{a_2(1-e_2)}{a_1(1+e_1)}>2.4\left[\max\left({M_{2}\over M_\star},{M_{1}\over M_\star}\right)\right]^{1/3}\left(a_2 \over a_1\right)^{1/2}+1.15\ ,
\end{equation}
which is applicable to planet-star mass ratios $10^{-4}-10^{-2}$
and mutual inclinations up to $40^{\circ}$. 
This criterion is essentially equivalent to that  of \citet{1995ApJ...455..640E}:
\begin{eqnarray}\label{ek}
\frac{a_2(1-e_2)}{a_1(1+e_1)}&>&1+3.7
  \left(M_{2}\over M_\star\right)^{1/3}+\frac{2.2}{1+\left(M_{2}/M_\star\right)^{-1/3}}+ \nonumber
  \\ 
&&1.4\left(M_{1} \over M_\star\right)^{1/3}\frac{\left(M_{2}/M_\star\right)^{-1/3}-1}{1+\left(M_{2}/M_\star\right)^{-1/3}}\ .
\end{eqnarray}
Systems that do not satisfy the condition 
Equation\ (\ref{pet+15}) and (\ref{ek}) are expected to be unstable leading to either ejections or collisions.
It can be shown by combining Equation\ (\ref{d+14-2})  and Equation\
(\ref{pet+15})  and taking the limit $e_1\rightarrow 1$ in this latter equation
that for a Jupiter mass perturber there are no stable configurations which
allow the formation of a WJ at $\lesssim 0.3\rm AU$.

Another often adopted stability criterion is that of \citet{2001MNRAS.321..398M},
\begin{equation}\label{MA01}
 \frac{a_2(1-e_2)}{a_1}> 2.8\left[\left(1+{M_{2}\over M_\star}\right)\frac{1+e_{\rm 2}}{(1-e_{\rm 2})^{1/2}} \right]^{2/5}~.
\end{equation}
Note that the \citet{2001MNRAS.321..398M}  criterion does not include a dependence on the inner planet orbital eccentricity,
and was derived for cases in which the mass ratio between the inner
and outer binary is not much different from unity. 
For these reasons, we consider Equations\ (\ref{pet+15}) and
(\ref{ek}) more accurate for the two planet systems we are considering. The results
of our simulations confirm this.

In Figure \ref{fig1} we compute the stability boundaries defined above
by adopting the observed 
orbital parameters of HJs and WJs with a detected companion. 
The full sample of planets we considered is presented in Table\ 1. We selected
 systems with two giant planets and that host a Jupiter
 planet with mass $M_1\sin i\geq 0.5 M_{\rm Jupiter}$ and semi-major axis $<1\rm AU$.
The left panel shows the inner Jupiter semi-major axis 
as a function of the critical inner-planet semi-major axis which would render the system unstable 
according to Equations\ (\ref{pet+15}) and (\ref{MA01}).
 At larger semi-major axis the system will be unstable and
any secular process leading to high-$e$ migration is likely to be suppressed.
Since a secular migration scenario requires the inner Jupiter to have initially an
extremely large eccentricity we take the limit  $e_1\rightarrow 1$  when evaluating  Equation\ (\ref{pet+15}).
Also, we added a factor 0.5 to right hand side of Equation\ (\ref{pet+15}),
which according to \citet{2015ApJ...808..120P} corresponds approximately to $95\%$ chance for a system to be unstable
over $10^8\rm years$ of evolution.

From the left panel of Figure \ref{fig1}  we see that most Jupiters at $0.6\leq a_1\leq0.8 \rm AU$
are close or above the stability boundary defined by the dashed line. 
Clearly this simple fact is difficult to reconcile with a high-$e$ migration model 
for these planets, suggesting that such systems  are unlikely to have experienced significant 
tidal migration from further out. Note that in  Figure \ref{fig1}   the stability
boundaries were computed using the minimum mass for the planets. If
the planet orbits were significantly tilted with respect to the line
of sight, the planet masses could be significantly larger which will
further  push the stability boundary towards smaller semi-major axes.

In the region $0.1\leq a_1\leq0.6 \rm AU$ the two adopted stability 
criteria start to give somewhat different values
for the limiting initial $a_1$ implied by our stability argument. 
According to Equation\ (\ref{pet+15}), 9 out of 10 WJs within 
$0.6\rm AU$ could not have migrated from $a_1\gtrsim 1\rm AU$ as
they would have been dynamically unstable otherwise.
According to Equation\ (\ref{MA01}) instead, 8 out of 10 WJs
within  $0.6\rm AU$  would be dynamically unstable at   $a_1\gtrsim 1\rm AU$
  suggesting that they could  not have migrated from these distances.
Two (HAT-P-13 and Kepler-424) of the four HJs with known  companion 
have very strong constraints on their maximum separation required by stability,
implying that if they formed through secular migration they 
must have been initially at $\lesssim 0.2\rm AU$.
 
A possibility is that WJs formed 
by interactions with a planetary companion and began tidal circularization interior to 1 AU after multiple scatterings. 
However, among the ten WJs detected within  $\lesssim 0.5\rm AU$ 
only two are above the stability boundary given by Equation\ (\ref{pet+15})
when setting $a_1=0.6\rm AU$ and taking the limit $e_1\rightarrow 1$; 
these planets are HD-37605c and HD-163607b.
There are also a few systems that at $a_1=0.6\rm AU$  would be 
classified as unstable according to Equation\ (\ref{pet+15})
but are just above the stability boundary defined by Equation\ (\ref{MA01}). These systems are 
HD-38529, HD-74156, HD-13908 and HD-168443. Further dynamical constraints on 
a possible high-$e$ migration scenario for some of these systems are
presented in the next sections.

In the right panel of Figure \ref{fig1}  we show the stability boundary in Equation\ (\ref{pet+15})
for our sample of two-planet systems. 
We also identify the region where $95\%$ of the systems according to Equation\ (\ref{pet+15})
would be unstable after $10^8\rm yr$ of evolution. 
Surprisingly, some of these planets appear to be well inside the dynamically unstable region.
As also noted by \citet{2015ApJ...808..120P}, however, the stability of
these  systems might be promoted
by mean-motion resonances (some of which are indicated in Figure \ref{fig1}).  In any case,
it is hard to imagine how these planets could have been transported  from further out 
through secular migration processes.

The stability analysis shown in  Figure \ref{fig1}  suggests that most WJs and HJs with 
observed companions cannot have migrated  to their current location via
tidal dissipation from $a_1\approx 1\rm AU$. This idea is further explored 
and supported by the analysis 
presented in the following sections.

In Table 1 we give the orbital parameters of observed two planet systems 
with an inner WJ or HJ and summarize the results of our stability analysis.
Importantly, and contrary to previous
work \citep{2014Sci...346..212D}, we note that our analysis  disfavors
a high migration origin for most WJs with a companion, including   those having
a finite orbital eccentricity ($e_1\gtrsim 0.2$).

\begin{figure}
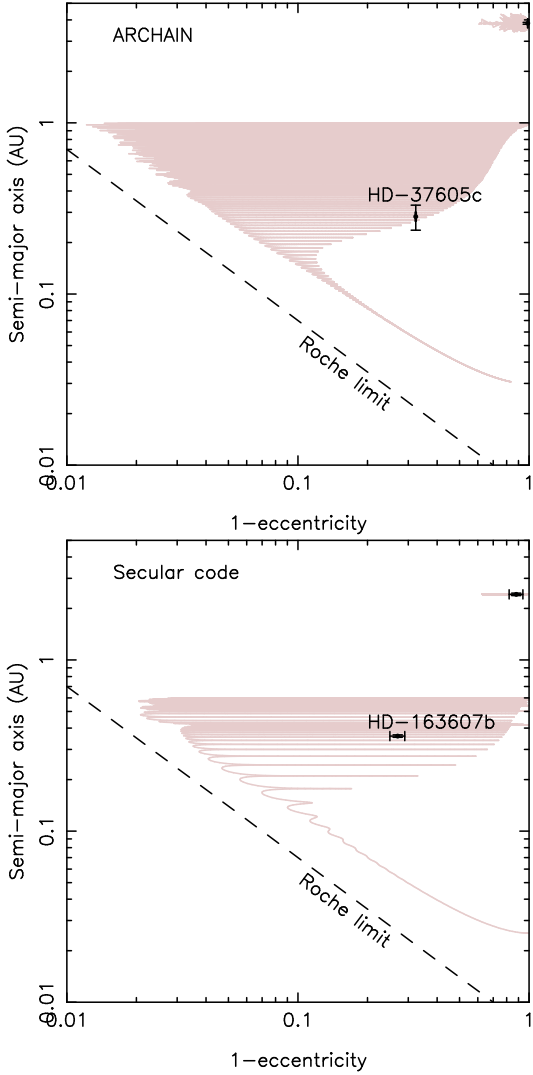

\centering
\includegraphics[width=2.8in,angle=270.]{Fig2a.eps}
\includegraphics[width=2.8in,angle=270.]{Fig2b.eps}
\caption{Evolution examples that lead to the formation of two planet systems
resembling  the observed systems HD-37605 and HD-163607. 
In the lower panel we evolved the system using the secular equations
of motion, in the upper panel we used the direct integrator ARCHAIN.
Dashed lines indicate the limit below which the planet 
will undergo Roche-lobe overflow. 
In both cases the systems are initially stable according to  Equation\
(\ref{pet+15}). The inner planet evolves to attain an orbit that is
consistent with the orbits of the observed planets. These two systems
represent therefore possible candidates for a secular migration origin, although
our stability analysis suggests that HD-163607b could not have
migrated from distances much larger than $\sim 0.6\ \rm AU$.
}\label{exmps}
\end{figure}

\begin{figure*}
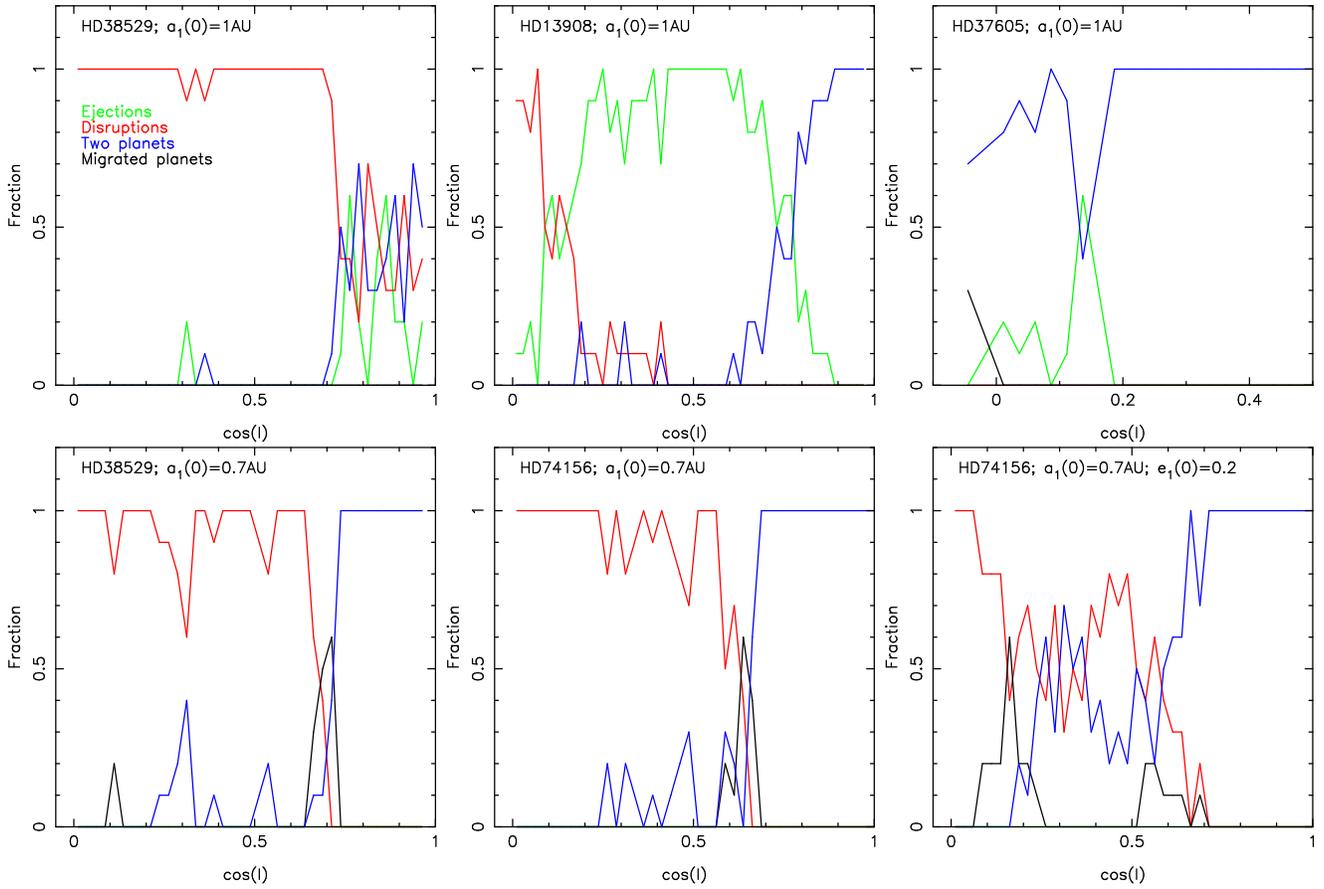

\centering
\includegraphics[width=2.3in,angle=270.]{Fig3a.eps}
\includegraphics[width=2.3in,angle=270.]{Fig3b.eps}
\includegraphics[width=2.3in,angle=270.]{Fig3c.eps}
\\
\includegraphics[width=2.3in,angle=270.]{Fig3d.eps}
\includegraphics[width=2.3in,angle=270.]{Fig3e.eps}
\includegraphics[width=2.3in,angle=270.]{Fig3f.eps}
\caption{
Results of the direct $N$-body integrations. In the upper panels the
inner planet was started at $a_1(0)=1\rm AU$, while in the lower panels
at $a_1(0)=0.7\rm AU$. 
The initial values of $a_2$, $e_2$, $M_2$ and
$M_1$ are given in Table 1 for each system. Here we set the
initial masses equal to the minimum mass given in the table. 
Note that no HJ and WJ is produced for configurations
that violate the stability criteria of Section\ \ref{s1}. When the
initial conditions are stable for large mutual inclinations (bottom panels), LK cycles
combined with tidal friction lead to the formation of HJs and WJs. Systems in which one of the two planets is ejected during
the simulation are indicated as ``Ejections'';  Systems in which the
inner planet had crossed its Roche limit are indicated as
``Disruptions''. ``Migrated planets'' are systems in which  at the end of the
simulation  the innermost planet was at $a_1\leq 0.5 \rm AU$. 
Systems that are stable and in which the innermost planet did not
experience significant tidal dissipation ($a_1> 0.5 \rm AU$) are
indicated  as ``Two planets''. 
}\label{fig-st} 
\end{figure*}

\begin{table*}
\caption{Observed orbital elements of detected systems comprising 
warm ($0.1 \leq a\leq 1\rm AU$) and hot  ($a < 0.1 \rm AU$)
Jupiters
and their close friend.
Only systems with two giant planets were considered.}
\begin{tabular}{llllllllll}
 \hline
System & $a_1$& $e_1$ &  $M_1 \sin i_1$ & $a_2$ & $e_2$  & $M_2 \sin i_2$ & $r_{\rm cr}$ &High-$e$\\ 
~&  (AU) &\phantom & ($M_{\rm Jupiter}$)  & (AU) & \phantom &($M_{\rm Jupiter}$) & (AU) & migration\\
\hline 
{\scriptsize HD82943} & $0.742 \pm 0.0129$ & $0.425 \pm 0.03$ & $1.59\pm 0.103$
 & $1.185\pm 0.022$ & $0.203\pm 0.065$ & $1.589\pm 0.097$& 0.50 &
                                                     $\times \times$ \\
{\scriptsize HD12661} & $0.838\pm 0.0177$ & $0.3768 \pm 0.0077$ & $2.34\pm 0.101$ &
$2.919 \pm 0.064$ & $0.031\pm 0.022$ & $1.949\pm 0.092$ & 1.19&
                                                     $\checkmark$\\
{\scriptsize HD169830} & $0.813\pm 0.0136$&$0.310\pm0.01$&$2.89\pm0.102$ 
&$3.60\pm0.35$& $0.330\pm0.02$&$4.06\pm0.35$& 0.87 &    
                                                     $\times$  \\
{\scriptsize HD207832} & $0.570\pm0.02$ & $0.13\pm0.05$ & $0.564\pm0.065$& 
$2.11\pm0.1$& $0.27\pm0.1$ & $0.73\pm0.161$ & 0.73&
                                                     $\times$  \\
{\scriptsize HD73526} & $0.647\pm0.011$ & $0.190\pm0.05$& $2.86\pm0.172$&
$1.028\pm0.0177$ & $0.140\pm 0.09$ & $2.42\pm 0.167$ & 0.43 &
                                                              $\times \times$  \\
{\scriptsize HD155358} & $0.627\pm0.0168$ & $0.170\pm0.03$ & $0.819\pm0.068$ &
$1.001\pm0.027$ & $0.16\pm0.1$ & $0.807\pm0.056$ & 0.47 &       
                                                          $\times \times$  \\
{\scriptsize HD202206} & $0.812\pm0.0164$ & $0.4350\pm 0.001$ & $16.82\pm 0.68$&
$2.490\pm0.055$ & $0.267\pm0.021$ & $2.33\pm 0.127$ & 0.65 & 
                                                             $\times \times$  \\
{\scriptsize HD60532} & $0.759 \pm 0.0176$ & $0.280 \pm 0.03$ & $1.035 \pm 0.069$ &
 $ 1.580\pm 0.04$ & $0.020 \pm 0.02$ & $2.46\pm 0.146$ & 0.73 &  
                                                                $\times \times$  \\ 
{\scriptsize HD134987} & $0.808\pm0.016$&$0.2330\pm0.002$ & $1.563\pm0.062$&
$5.83\pm0.33$ & $0.120\pm 0.02$ & $0.805\pm 0.046$ & 1.93 & 
                                                            $\checkmark$  \\
{\scriptsize HD37605} & $0.283\pm0.047$ & $0.6767\pm0.0019$ & $2.80\pm 0.93$&
$3.82\pm0.64$ & $0.013\pm 0.013$ & $3.4\pm1.12$ & 1.34 & 
                                                         $\checkmark$ \\
{\scriptsize HD163607} & $0.3592\pm 0.006$ & $0.730\pm 0.02$ & $0.769 \pm 0.041$ &
$2.418 \pm0.041$ & $0.120\pm 006$ & $2.29\pm 0.108$ & 0.76 &
                                                             $\times$ \\
{\scriptsize HD147018} & $0.2389 \pm 0.004$ & $0.486 \pm 0.0081$ &$ 2.127\pm 0.076 $& 
$1.923 \pm 0.039 $& $0.133 \pm 0.011 $& $6.59 \pm 0.29$ & 0.62 &
                                                                 $\times$ \\
{\scriptsize HD74156} & $0.2915 \pm 0.0049$ & $0.630 \pm 0.01$ & $1.773 \pm 0.09$ &
$3.900\pm0.067$  &  $0.380\pm 0.02$ & $8.25\pm 0.36$ & 0.86& 
                                                            $\times$ \\
{\scriptsize HD13908} & $0.1538 \pm 0.0026$ & $0.046 \pm 0.022$ & $0.865 \pm 0.035$ &
$2.034\pm 0.042$ & $0.120\pm 0.02$ & $5.13\pm 0.25$& 0.64 & 
                                                    $\times$~/  \\
{\scriptsize HD168443} & $0.2939 \pm 0.0049$ & $0.529 \pm 0.024$ & $7.70 \pm 0.29$ &
 $2.853 \pm 0.048$ & $0.2113 \pm 0.0017$ & $17.39\pm 0.58 $ & 0.80 & 
                                                                   $\times$   \\
{\scriptsize HD159243} & $0.1104 \pm 0.0018$& $0.020 \pm 0.018$ &$1.130\pm 0.05$ &
$0.805\pm 0.0171$ & $0.075\pm 0.05$  & $1.90\pm 0.13$ & 0.27 &
                                                               $\times$~/ \\
{\scriptsize HD38529} & $0.1272\pm 0.0021$ & $0.244 \pm 0.028$ & $0.803\pm0.033$ &
$3.600\pm0.06$ & $0.3551\pm0.0074$ & $12.26\pm 0.42$ & 0.83 &
                                                              $\times$~/\\
{\scriptsize HD9446} & $0.1892\pm0.0063$ & $0.200\pm0.027$ & $0.699\pm0.065$ &
$0.654\pm0.022$ & $0.060\pm 0.06$ & $1.82\pm0.172$ & 0.27 &
                                                 $\times$~/ \\
{\scriptsize TYC-1422} & \\
{\scriptsize-614-1} & $0.689\pm 0.036$ & $0.06\pm 0.02$ & $2.5\pm 0.4$ &
$1.396\pm0.073$ &$0.048 \pm 0.014$ & $10 \pm 1$ & 0.51&
                                                        $\times \times$\\
{\scriptsize K-432} & $0.301\pm 0.065$ & $0.5134\pm 0.0089$ & $5.5\pm 2.4$ & 
$1.18\pm0.25$ & $0.498\pm 0.059$ & $2.4\pm 1.04$ & 0.21& 
                                                         $\times \times$\\
\hline
{\scriptsize K-424} & $0.04365\pm 0.00078$  & $0.002\pm 0.071$ & $1.034\pm0.099$ &
$0.724\pm 0.0137$ & $0.319\pm 0.081$ & $7.04\pm0.58$& 0.18 & 
                                                             $\times$\\
{\scriptsize HAT-P-13} & $0.04269\pm0.00087$ & $0.0133\pm0.0041$ 
&$0.851\pm 0.035$ & $1.226\pm0.025$ & $0.6616\pm 0.0054$ 
& $14.27\pm 0.69$ & 0.14 & 
                           $\times$\\
{\scriptsize HD217107} & $0.0750\pm 0.00125$ & $0.1267 \pm 0.0052$ & $1.401\pm 0.048$ &$5.33\pm0.2$&
$0.517\pm0.033$ & $2.62\pm 0.15$ & 0.91 & 
                                          $\checkmark$ \\
{\scriptsize HD187123} & $0.04209\pm 0.0007$ & $0.0103\pm 0.0059$ & 
$0.510\pm 0.0173$ & $4.83\pm0.37$ & $0.252\pm 0.033$ & $1.94\pm
                                                       0.152$& 1.30&
  $\checkmark$\\ 
\hline \hline
\end{tabular}
{Orbital parameters 
of known WJs and HJs with a detected Jupiter companion. We selected
 systems with two giant planets,  hosting a Jupiter
 planet with mass $M_1\sin i\geq 0.5M_{\rm Jupiter}$ and semi-major axis $<1\rm AU$.
The
value of  $r_{\rm cr}$ is the maximum value of the inner planet
semi-major axis above which the two planet system will be dynamically
unstable, which we computed as  the maximum value between the two stability boundaries
obtained via Equation\ (\ref{pet+15}) and Equation\
(\ref{MA01}). 
The stability limit imposed by Equation\ (\ref{pet+15}) 
was computed taking the limit $e_1 \rightarrow 1$ and adding a factor 0.5 to the
right hand side
 which corresponds to $95\%$ chance for a system to be unstable over $10^8$ years.
Systems with $ r_{\rm cr}\lesssim 1\rm AU$ ($ r_{\rm cr}\lesssim a_1$) are indicated with a
$\times$ ($\times\times$) symbol in the last column. 
Our stability analysis disfavors a high-migration scenario for the formation of these
systems.
Systems with a low eccentricity, $e_1\leq 0.3$, at $0.1\leq a_1 \leq
0.3\rm AU$ are indicated with a $/$ symbol. Our analysis 
of Section\ \ref{dmcs} disfavors a high-migration scenario for these
systems as well.   Systems for which our 
study does not rule out a secular migration origin are indicated with a 
$\checkmark$ symbol.  The observations reported 
at http://exoplanets.org \citep{2011PASP..123..412W}  include data  from
\citet{2004AA...415..391M,2005AA...440..751C,2006ApJ...647..594T,2008AA...491..883D,2009ApJ...693.1084W,2010AA...511A..45S,2010AA...513A..69H,2010ApJ...718..575W,2010MNRAS.403.1703J,2011ApJ...743..162P,2011ApJ...727..117M,2012ApJ...761...46W,2012ApJ...756...91H,2012ApJ...744....4G,2012ApJ...749...39R,2013ApJ...777..101T,2014ApJ...795..151E,2014AA...563A..22M,2015AA...573A..36N,2015ApJ...803...49Q}.\\
}\label{t1}
\end{table*}

\section{Numerical methods and test cases}
In what follows we 
 study the evolution of Jupiter like planets around a solar like star
induced by the gravitational interaction with an outer Jupiter companion.
Our goal is to put constraints on the
origin of some of the observed Jupiters within $1\rm AU$ focusing
mostly on planets within the period valley that have  a close companion.
In order to do so, we use two numerical approaches: 
(i)  direct $N$-body integrations of the equations of motion 
and (ii) integrations of the orbit averaged secular equations 
of motion. 
In both cases we included terms to the equations of motion that account for Schwarzschild precession, apsidal precession due to 
tidal bulges and  terms which account for tidal dissipation.

The direct  integrations presented  
 below were performed using  ARCHAIN \citep{2008AJ....135.2398M}. 
ARCHAIN employs an algorithmically regularized chain structure and the time-transformed leapfrog scheme 
which allow to integrate the evolution of the motion of arbitrarily tight binaries with
arbitrarily mass ratio with extremely high precision. The code
includes post-Newtonian (PN) non-dissipative 1PN, 2PN and dissipative
2.5PN corrections to all pair-forces. To these we
also added terms that account for precession induced by  tidal bulges as
well as tidal dissipation.
Velocity-dependent forces were implemented using the 
 generalized mid-point method described in \citet{2006MNRAS.372..219M}.
The tidal perturbation force was set equal to \citep{1981AA....99..126H}
\begin{equation}\label{tides} 
{\bf F}=-G \frac{M_\star M_1}{r^2}
 \left\{
3\frac{M_\star}{M_1}  \left( \frac{R}{r}\right)^5k 
\left( 
1+3\frac{\dot{r}}{r}\tau 
\right) \hat{r} 
\right\} \ ,
\end{equation}
where $M_1$ is the mass of the planet and
$R$ its radius hereafter  set equal
to one Jupiter radius, 
$k$ (set to 0.28) is the apsidal
motion constant and $\tau$ is the $constant$ time-lag factor.
Hereafter we use $M_\star=1M_{\odot}$.

 The secular integrations performed in this paper make use of the standard octupole level 
secular equations of motion of \citet[][their Eq. 11-17]{2002ApJ...578..775B}, including 
terms accounting for relativistic precession. We added terms that describe 
 apsidal  precession induced by tidal bulges and tidal friction.
The perturbing acceleration\ (\ref{tides}) causes a slow change of
the orbital parameters.
Following \citet{2012arXiv1209.5724S}, 
in the limit of high $e$, 
the orbit average change 
 rate corresponding to Equation\ (\ref{tides}) is
\begin{equation}\label{td1}
\frac{\dot{a}}{a}=-\frac{4059}{320t_D} \sqrt{\frac{a_F}{a}}\ ,
\end{equation}
\begin{equation}\label{td2}
\frac{\dot{e}}{e}=\frac{\dot{a}}{a} \frac{\left(1-e^2\right)}{2e^2}\  ,
\end{equation}
and
\begin{equation}\label{td3}
\dot{\omega}=\frac{15\left[G(M_\star)\right]^{1/2}}{8a_1^{13/2}}\frac{8+12e_1^2+e_1^4}{\left(1-e_1^2\right)^5}  
\frac{M_\star}{M_1}kR^5,
\end{equation}
where $a_F=a\left(1-e^2 \right)$, and
 $t_D$ is the characteristic time for tidal dissipation:
\begin{equation}
t_D=\frac{M_1a_F^8}{6k\tau G M_\star^2R^5} \ .
\end{equation}
In our simulations we neglected any additional precession induced by
the stellar host rotational bulge. For the cases considered here
we find in fact that  the precession due to tidal bulges is dominant
and rotational bulges
become important only for rapidly
rotating stars with spin period less than $\approx 1\rm day$.
Finally we  assume that the inner planet was disrupted by its host
star tidal field if it crossed the Roche limit:
\begin{equation}
a_1(1-e_1)\leq 0.01 \frac{R}{R_{\rm
    Jupiter}}\left(\frac{M_\star}{M_{\odot}} \frac{M_{\rm Jupiter}}{M_{1}} \right)^{1/3}\ \rm AU .
\end{equation}

 The secular integrations base on two 
  two levels of approximation being implemented: (i) double-orbit averaging, (ii) perturbation up octupole-order.
More specifically, 
the orbit average approximation, on which the 
\citet{2002ApJ...578..775B} treatment is based on, 
breaks down  if\ \citep{2014ApJ...781...45A}:
\begin{equation}
\frac{a_2(1-e_2)}{a_1}\lesssim {0.2} \left({M_\odot \over M_\star} {M_2\over M_{\rm Jupiter} }\right)^{1/3}\
 \left({a_1\over a_{\rm cr}}0.1\right)^{1/6} \ .
\end{equation}
For values of ${a_2}/{a_1}$ smaller than the one given by this last equation 
the orbital angular momentum of the inner planet can undergo oscillations 
on a timescale shorter than its  orbital period.
Clearly, this condition is never met for the two-planet systems considered here
unless the orbits of the planets are crossing.
Mean-motion resonances are also neglected in our secular integrations.
We note that for $a_2/a_1\approx 3$, their effect might not be fully negligible. The potential associated with $q$-th order mean-motion resonances has terms with amplitude $\sim e^q$ and since the eccentricities are order unity (especially for migrating planets) their effect might be significant.
Moreover, even if the effect of mean-motion resonances is negligible, then the octupole-level expansion might not  resolve the behavior of the $N$-body properly for low $a_2/a_1$
as more terms in the expansion might be needed. 
Although we caution on the simplifications implemented in our treatment
the results of secular and direct integrations  were compared  for a number of initial conditions
and found  to give in general consistent results.

In Figure\ \ref{exmps} we show two example cases. 
The top panel shows a three body integration with initial conditions
representing a possible progenitor for HD-37605c and initial mutual inclination $I=94^\circ$. 
The observed system consists of an eccentric ($e_1=0.68$) WJ at $a_1=0.28\rm AU$ and minimum mass  $M_1\sin i=2.8M_{\rm Jupiter}$,
with a companion at  $a_1=3.8\rm AU$ and minimum mass  $M_2\sin i=3.4M_{\rm Jupiter}$.
Accordingly to our analysis of Section\ \ref{s1}, this is the only two
planet system hosting a WJ within $\leq 0.5\rm AU$
that would be dynamically stable if we were to place the inner planet at $1\rm AU$ on a highly eccentric orbit.
Thus, HD-37605c is a possible candidate for a secular migration origin.
The example in the upper panel of Figure\ \ref{exmps} shows that this secular migration scenario is indeed a possibility for such a system.
The inner planet starts at $a_1(0)=1\rm AU$
and evolves to become a HJ. During this transition
the inner planet orbital semi-major axis and eccentricity take values that are consistent with the observed orbit of HD-37605c.
Note that the amount of time the system spends in this region of parameter space depends on the efficiency of tidal dissipation
which in turns is regulated by the poorly constrained value of $\tau$ in Equation\ (\ref{tides}).
However, the characteristic shape of the envelope within which the 
planet orbit evolves does not depend significantly on the assumed
value of $\tau$ (See also Section\ \ref{nbd}). 

For the bottom panel of Figure\ \ref{exmps} we show a secular integration
of a two planet system with initial conditions that 
resemble the observed system HD-163607  and mutual inclination $I=84^\circ$. 
In this case we start the inner planet at $a_1(0)=0.6\rm AU$. 
The observed system consists of an eccentric ($e_1=0.73$) WJ at $a_1=0.36\rm AU$ and minimum mass  $M_1\sin i=0.77M_{\rm Jupiter}$,
with a companion at  $a_1=2.4\rm AU$ and minimum mass  $M_2\sin i=2.3M_{\rm Jupiter}$. Even in this case
the inner planet evolves through a region of parameter space which is consistent with the observed orbit
of the planet HD-163607b.

\section{$N$-body simulations: near the edge of stability}\label{nbd}
In this section we consider the evolution of 2 planet systems that are close to
the stability boundary defined by Equation\ (\ref{pet+15}).
In particular we focus on systems with properties that  resemble
those of the two planet systems HD-38529, HD-74156 and HD-13908. 
(We specifically analyzed  the stability of these three systems because 
their value of $r_{\rm cr}$ in Table 1  is just below $\rm 1 AU$.
Thus, given the uncertainty in the adopted stability criteria, it is unclear
whether these systems will be actually unstable at $a_1\approx 1\rm AU$.)
The results of these simulations are used to validate the
stability criteria adopted above and our argument that
systems  which are dynamically unstable according to these
criteria do not lead to tidal migration but
rather to planet ejections and collisions with the stellar host.

We run 1200 direct integrations, 200 per panel in Figure\ \ref{fig-st}. 
The initial mutual inclination between the inner and outer planet orbits, $\cos(I)$,
was sampled uniformly
 between $0$ and $1$.
 The initial inner and outer planet argument of periapsis $\omega_1$ and
$\omega_2$ and the longitude of the ascending nodes
$\Omega_1$ and $\Omega_2$
were chosen randomly between $0$ and $2\pi$.
The outer planet initial eccentricity and semi-major axis were set equal to the observed values
while the inner planet eccentricity was initially set to a fixed value
(0.01 and 0.2). We take the mass of the planets to be equal to the minimum
mass  as inferred from observations. 
We  set $\tau=66\rm sec$, and evolved 
each  system for a maximum time  of
$10^8\rm yr$.
In addition we run ten retrograde configurations, $\cos(I)$ in the range $(-0.1-0)$, 
for initial conditions 
corresponding to HD37605.
By comparing a number of orbit-average integrations in which we
 adopted  different values of $\tau$,
we found that the results of these integrations  can be rescaled 
using
\begin{equation} \label{scal}
t'\rightarrow t \times \frac{\tau}{\tau'} ,
\end{equation}
so that evolving a system for $10^8\rm yr$ with  $\tau=66\rm sec$
would be at a good approximation equivalent to evolve the same system for $10^{10}\rm yr$
with  $\tau=0.66\rm sec$.
This latter value of $\tau$ is large enough to
allow the formation of HJs at $\lesssim 0.1\rm AU$ in
$10^{10}\rm yr$ \citep{2012arXiv1209.5724S}.
We caution that although the scaling of Equation\ (\ref{scal}) is almost exact for 
stable systems, it might be an oversimplification  near the region of
instability given that our direct simulations
cannot identify whether a system will be unstable on timescales longer
than $10^8\rm yr$. In addition to this, the outer planet mass
could be larger than the adopted value which will render the system
even more susceptible to dynamical instabilities. 
It is likely therefore that in our analysis 
we are overestimating  the number of stable systems.

Figure \ref{fig-st} displays the results of the direct integrations.
In the upper panels the inner planet semi-major axis is initially $a_1(0)=1\rm AU$.
In these cases most configurations
are unstable leading to planet disruptions  (red curves)
or ejection of one of the planets (green curves).
We also checked for any collision between the two planets 
but did not find any. We calculate the number of ``migrated'' planets as those that have
reached  within $a_1\leq 0.5\rm AU$ at the end of the integration.
As expected, for configurations that are unstable according to Equation\
(\ref{pet+15})  no migrating planet was formed. 
Three planets had $a_1\leq 0.5\rm AU$  for the initial conditions
corresponding to HD37605c,
however, only for retrograde configurations. This is a consequence of the
back-reaction torque of the inner planet  on the outer orbit which shifts the 
initial critical inclination at which the maximum possible $e$ is
attained at $\geq 90^\circ$.
Note that a more massive perturber will reduce this effect and  allow the formation of WJs
and HJs also for prograde configurations.

In the bottom panel of Figure \ref{fig-st} we set $a_1(0)=0.7\rm AU$.
These configurations are stable according to our stability criteria, although they
are near the extreme of inequality (\ref{MA01}).
 At large inclinations, $\cos (I) \lesssim 0.5$,
the inner planet orbit becomes extremely eccentric so that in most cases the 
planet collides with the star. This is expected since in all cases
considered here the 
maximum eccentricity attained by the inner planet
  is not limited by precession due to tidal  bulges which becomes a limiting factor for the maximum $e_1$ 
 only  at $a_2/a_1\gtrsim 4$.  
Nevertheless, a few planets managed to migrate  within $0.5\rm AU$ for a  mutual 
inclination that lies initially  near or above the LK critical angle, $\cos (I)
\approx 0.65$. In these cases, the inner planets
attain an eccentricity that is large enough to allow for efficient tidal dissipation, 
but, due to the mild initial inclinations, never high enough to cause the disruption of
the planet.
 For  initial inclinations smaller than $\cos (I)
\approx 0.65$, the inner
planet eccentricity cannot be excited  to high values so that the system
is more stable and the orbital parameters of the planets remain
essentially unchanged during the evolution. 

In conclusion,  our direct integrations show that LK induced
migration is unlikely to occur for unstable systems while it can 
lead to the formation of hot and warm Jupiters for systems that are
just below  the stability boundaries, although, as shown next, these
are likely to be rare.

\begin{figure}
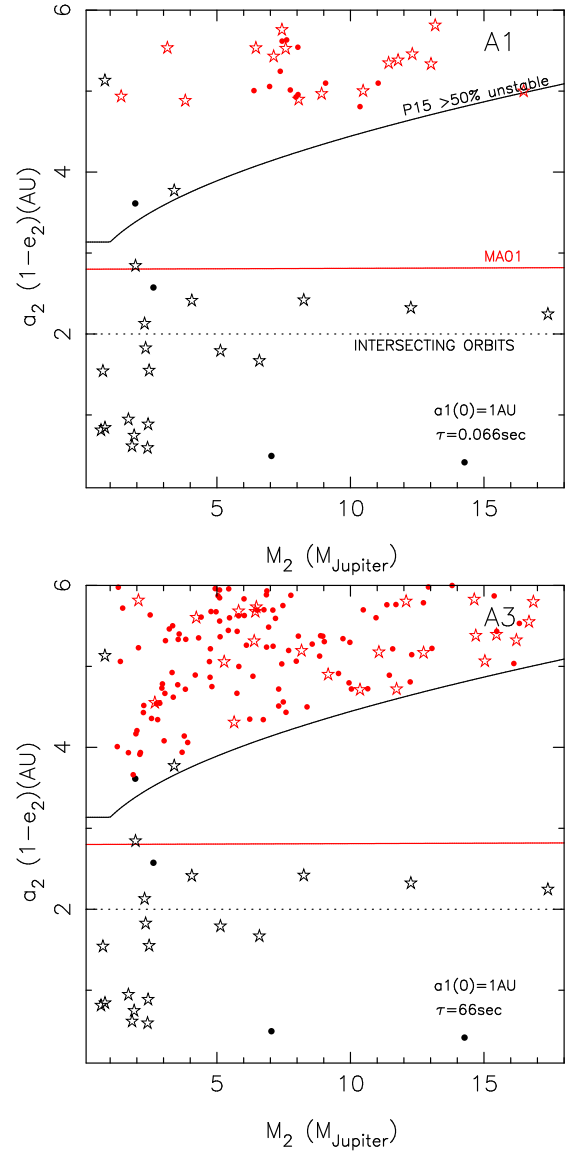

\centering
\includegraphics[width=3.in,angle=270.]{Fig4a.eps}
\includegraphics[width=3.in,angle=270.]{Fig4b.eps}
\caption{
Solid lines show regions of stability. Any proto-HJ/WJ with 
$a_1=1\rm AU$, $e_1=1$, and $M\sin I \geq 0.5 M_{\rm Jupiter}$ would be unstable
according to Equation (\ref{MA01}) if its external companion lay below the horizontal red line (labelled MA01).
The other solid line corresponds to Equation (\ref{pet+15}).  
Below the dashed line the two planets would be on intersecting orbits.
Simulations results are shown for models A1 and A3 (see Table 2).
Red stars indicate WJs, i.e., those systems in which the inner
planet has experienced significant tidal dissipation and migrated
within $0.9\rm AU$ but has not
evolved inside $0.1\rm AU$. 
Red points  are systems that have
formed a HJ ($a_1< 0.1\rm AU$) at the end of the integration. 
Black symbols correspond to the observed systems of Table 1.
Black stars are WJs ($a_1=0.1-1\rm AU$); black  points  are
HJs ($a_1<0.1\rm AU$).
} \label{hem-res}
\end{figure}

\begin{figure}
\centering
\includegraphics[width=2.5in,angle=0.]{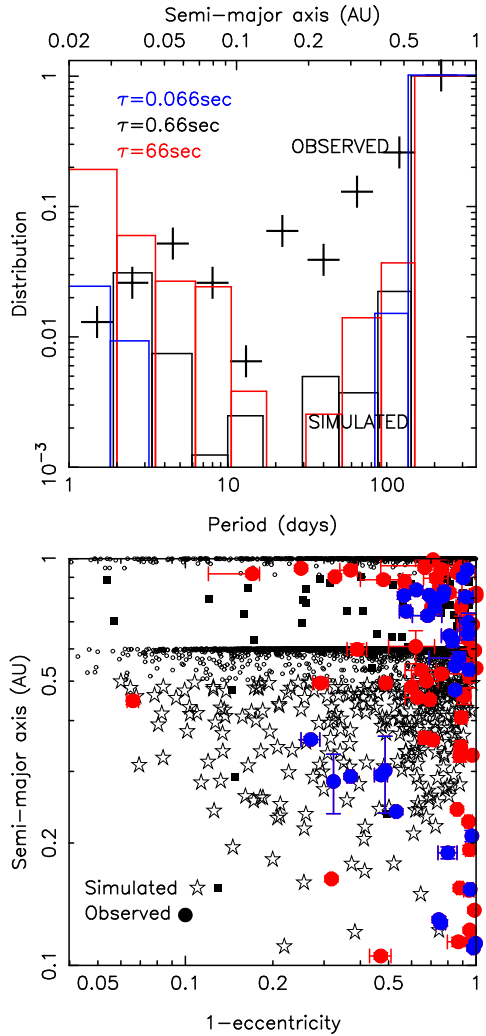}
\caption{
Upper panel: semi-major axis distribution of the planets
from our simulations (histogram) compared to the observed distribution from
\citet{2015arXiv151100643S} (crossed symbols). 
The simulated distributions are from the C1-c, C2-c and C3-c models
that have $a_1(0)=0.6\rm AU$
(see Table 2).
Other models that have $a_1(0)=0.6\rm AU$ were found to produce
similar distributions. 
The observed and simulated distributions have been normalized such to have the
same value at  periods $P\gtrsim 200\rm days$. Note how any of our models
greatly underestimates the number of  Jupiters in the WJ zone.
Lower panel: $a$ vs $e$ distribution of the inner planets for 
systems that were evolved forward in time using the secular equations of
motion. Small open circles
are systems that have not experienced 
significant inward migration. Filled square symbols (star symbols) correspond to
planets with $a_1(0)=1\rm AU$ ($a_1(0)=0.6\rm AU$) and that migrated within
$0.9\rm AU$ ($0.5\rm AU$) by the end of the simulation.
Note that models with $a_1(0)=1\rm AU$ produce almost no
WJs inside $0.5\rm AU$.
These distributions are
compared to observational data which are represented by the filled
circles.  Data points corresponding to Jupiters
with a known Jupiter companion are in blue. In the lower panel data
are from exoplanets.org.
}\label{a-e-distr}
\end{figure}

\section{Orbit average treatment: migrating planets and their orbital distribution}\label{hem}

\begin{table*}
\centering
\caption{Results of secular integrations}
\begin{tabular}{llllllllll}
 \hline
  Model& Stability &$a_1(0)$& $e_1(0)$  & $e_2(0)$  & $\tau$ & HJs & WJs
  & Non-migrating & Disrupted  \\
\phantom & criterion  &$(\rm AU$)&  \phantom & \phantom  & $(\rm sec)$ & $\%$ & $\%$  & $\%$ & $\%$ \\
\hline 
A1     & P15 &$1\rm AU$ & $0.1$ & $0.2$ & $0.066$ & 0.77 & 1.02 & 70.7 & 27.5\\
A2     & - &$1\rm AU$ & $0.1$ & $0.2$ & $0.66$ & 1.22 & 0.90 & 64.4 &  33.5 \\
A3     & -&$1\rm AU$ & $0.1$ & $0.2$ & $66$ & 10.3 & 1.54 & 59.3 & 28.9 \\
B1     & -&$0.6\rm AU$ & $0.1$ & $0.2$ & $0.066$ & 1.73 & 0.70 & 66.4 &  31.2 \\
B2     & -&$0.6\rm AU$ & $0.1$ & $0.2$ & $0.66$ & 4.41 & 0.51 & 65.1 &  29.9 \\
B3     & -&$0.6\rm AU$ & $0.1$ & $0.2$ & $66$  & 21.8 & 0.83 & 56.9 & 20.5\\
C1     & MA01 &$0.6\rm AU$ & Rayleigh & Rayleigh  & $0.066$ & 3.01 & 0.32 & 68.5 & 28.2\\
C2     & -&$0.6\rm AU$ & Rayleigh & Rayleigh  & $0.66$ & 4.73 & 1.09 & 67.1&  27.1 \\
C3     & -&$0.6\rm AU$ & Rayleigh & Rayleigh  & $66$ & 19.7 & 2.05 & 60.4 & 17.9\\
C1-c  & -& $0.6\rm AU$ & Rayleigh & Rayleigh  & $0.066$ & 1.86 & 0.58 & 79.0 &  18.6 \\
C2-c  & -&$0.6\rm AU$ & Rayleigh & Rayleigh  & $0.66$ & 4.41 & 0.96 & 74.3 &  20.3 \\
C3-c  & -&$0.6\rm AU$ & Rayleigh & Rayleigh  & $66$ & 16.1 & 1.73 & 67.2 &  15.0 \\
\hline \hline 
\end{tabular}\\
{We define here as HJs those planets that   by the end of the simulation
  have migrated within $0.1\rm AU$; WJs are defined as those planets that 
are in the semi-major axis range $0.1-0.9\rm AU$ (for $a_1(0)=1\rm AU$) or
  $0.1-0.5\rm AU$  (for $a_1(0)=0.6\rm AU$). 
Non-migrating planets have final
semi-major axis $a_1\geq 0.9\rm AU$ (for $a_1(0)=1\rm AU$) or
  $a_1\geq  0.5\rm AU$  (for $a_1(0)=0.6\rm AU$);
  ``disrupted'' systems are those in which the inner planet had
crossed its Roche limit. In the first models we run the integrations up
  to a maximum time of $10\rm Gyr$. In the last three models (C1-c,
  C2-c, C3-c) the final integration time was chosen randomly between 
$0$ and $10\rm Gyr$. This latter choice is to simulate a scenario in
which the planets have formed continuously over the last $10 \rm Gyr$.
In order to account for the fact that we have only selected systems
with an inclination $50^\circ$,
 we have reduced the number of forming HJs and WJs (and disrupted) by
 $\cos \ 50^\circ$, and then used that number in the fraction.
}\label{tb2}
\end{table*}

In this Section we run suites of LK simulations, in which systems
are initizialized with two Jupiter mass planets at high mutual
inclination. 
By comparing the ratio of WJs to cold Jupiters (CJs)  found in
the simulations to that ratio as observed we put constraints on the 
fraction of observed WJs that  are likely to have undergone high-$e$ LK migration.

In our simulations we set the mass of the inner planet to $M_1=1\ M_{\rm Jupiter}$
and sample the mass of the outer perturber uniformly
in the range  $1 \leq M_2\leq 17\ M_{\rm Jupiter}$. These latter values correspond approximately
to the extremes of the mass distribution of the observed Jupiter
companions to HJs and WJs. 
We assume the orbits to be prograde and
sample the mutual inclination from a uniform distribution
in  $\cos (I)$. We take the mutual inclination in the range $50$
to $90$ degrees. The high mutual orbital inclination is required in
order for the inner  planet to reach high eccentricities.
We adopt a small initial eccentricity for the inner planet $e_1=0.1$
and take $e_2=0.2$ for the outer planet. This latter value is close to 
the median of the eccentricity distribution for the outer companion of
observed HJs and WJs.
The outer planet semi-major axis is sampled uniformly within $a_2\leq
8\rm AU$ with the lower limit set such that the system satisfied the
stability condition of  Equation\ (\ref{pet+15}) in the limit $e\rightarrow 1$.
We considered two values of the inner planet semi-major axis: $a_1=1\rm AU$, $a_1=0.6\rm AU$.
For each value of $a_1$ we considered three values of the time-lag 
constant $\tau=0.066$, $0.66$
and  $\tau=66\rm sec$, which
correspond to a tidal quality factor $Q$ of $\approx 10^6$, $10^5$ and $10^3$  respectively.
Thus, we evolved 6 sets of initial conditions. For each
set we performed a total of 1000 random realizations 
integrating them up to a  final
integration time of $10\rm Gyr$.

 In addition to the secular integrations described above we run 
3  sets of initial conditions (corresponding to the three
values of $\tau$) where we used a Rayleigh
distribution  for both $e_1$ and $e_2$ 
with a mean eccentricity of 0.175 \citep{2011ApJS..197....1M}, and this time sampling $a_2$
 such that the system satisfied the stability condition of Equation
 (\ref{MA01}).  In these latter simulations (models C in Table\ \ref{tb2}), the planets  
can have initially a lower  $a_2/a_1$ ratio which favors the formation of WJs 
as discussed in Section\ \ref{s1}.
As a consequence of this, such models are expected to produce more 
 WJs than if we were to select the initial conditions  based on inequality\ (\ref{pet+15}),
 however we caution that they also contain more systems that are
near the orbit crossing when $e_1\approx 1$ and for which
 the orbit-averaged treatment we use is less accurate.

Table 2 gives the initial setup of the numerical integrations
and summarizes the main results of our simulations,
giving the fraction of systems that lead to the specified outcomes.
In order to take into account the fact that we are only simulating systems
with mutual inclinations larger than $50^\circ$ 
we reduced the number of forming HJs and WJs (and disrupted) by $\cos\ 50^\circ$, and then used that number in the fraction.
We find that a high-$e$ LK migration scenario is more efficient at
producing HJs than WJs.
The fraction of systems that result in
the formation of WJs is  $\approx 1\%$ of the total  and it is roughly constant, 
having little variation with the initial conditions and tidal dissipation strength.
 The fact that the number 
of $WJs$ is not very sensitive to $\tau$  is because
increasing $\tau$  will result in a larger  number of planets 
migrating inside the period valley, but at the same time
also in a larger number of planets leaving it.
According to our results, if WJs are migrating planets, then for a homogeneous
sample of planets we 
we would expect that planets with detected outer companions would more often be HJ’s than WJ’s.
This expectation  appears to be at odds  with
what is observed. Restricting our analysis to the subset of known
extrasolar planets discovered by radial-velocity surveys, only 2 of
the Jupiter mass planets ($M_1\sin i\ge 0.5\ M_{\rm Jupiter}$) at $\leq 0.1\rm \ AU$ have a detected
outer companion within $\lesssim 5 \rm AU$ distance from their host star,  
while 20 Jupiters mass planets at $0.1\rm
AU\leq a_1\leq 1 \rm \ AU $ do.  This suggests that HJs and WJs belong to two distinct populations
of planets which likely  originated through different processes.

Figure \ref{hem-res}  shows 
the properties of outer companions
to the observed HJs (black points) and WJs (black stars).
The black and red curves show the value of $a_2$
satisfying the condition Equations\ (\ref{pet+15}) and (\ref{MA01})
respectively.
Below these lines  WJs are unlikely to form as the system is 
unstable to either collision or ejection
of one of the planets. 
As done previously in Section \ref{s1}  we computed the stability
limit from Equations\ (\ref{pet+15}) by taking the
relevant  limit  $e_1\rightarrow 1$.
In Figure \ref{hem-res} the
relative position of the black  points to the stability lines show that
only four of the twenty-four observed  systems would be dynamically
stable if the inner planet had $a_1=1\rm AU$, and fourteen of them
would be on intersecting orbits.
Figure \ref{hem-res}  also displays  the results from models A1 and A3 of
Table 2,
showing the systems that form HJs (red points) and
those that form WJs (red stars).
These latter being defined as those systems 
in which the innermost planet had experienced significant tidal
dissipation and migrated inside the semi-major axis range  $0.1-0.9\rm AU$.
From this figure we see again that the number of migrated planets that have formed a HJ increases
significantly  when increasing the time-lag factor $\tau$,
while the number of migrating planets in the period valley 
(red stars) increases but
only slightly. Accordingly,  the number of HJs
formed in our models is typically equal or larger than the number
of migrating WJs.

In the upper panel of Figure\ \ref{a-e-distr} we show the period
 distribution of migrating planets and compared this to the
(intrinsic) observed distribution from
\citet{2015arXiv151100643S}.
Such comparison shows that the
semi-major axis distribution of the planets in our simulations does
not provide a good match to the observed period distribution of gas
giants in the period valley.  
Given the observed number of Jupiters at $P\gtrsim 100\rm days$, the models underpredict
the number of migrating planets below this period by at least one
order of magnitude. 
We conclude that a
high-$e$ migration mechanism can be responsible for  less than $10\%$ of all 
gas giants with orbital periods in the range $10-100\rm days$.

We note that our simulations are restricted to systems with inner
planets initially at $\leq \rm 1 AU$, accompanied by a outer perturber within
 $\leq \rm 8 AU$. 
A significant contribution to the WJ population from planets migrating
from  $\gg \rm 1 AU$ and having companions at  $\gg \rm 8 AU$ is
unlikely. In fact, as also noted in Section \ref{s1},
WJs cannot form  for Jupiter mass perturbers if $a_2$ is larger than a few AUs.

Finally, we note that similar to previous studies \citep[e.g.,][]{2015arXiv151008918A}, our model distributions do not match  the period distribution of
HJs ($<10\rm\ d$), producing  too many planets at short 
orbital periods ($\sim1\rm\ d$).   
Such discrepancy might 
depend on the choice we made for various parameters, most importantly on the 
adopted value of the planet radius  which determines the final semi-major axis
 of the HJ \citep{2011ApJ...735..109W}.

A secular migration model for WJs fails at explaining some additional
features of the observed
orbital distribution.  
First, in only three of the 3000 models that were  started at $a_1(0)=1\rm AU$ 
the inner planet was found  in the region $0.1-0.5\rm
AU$ by the end of the simulation. 
This is  in contrast with the 
abundance of observed Jovian planets at these radii
and casts further doubts on a possible migratory origin from $\gtrsim 1\rm AU$ for
these systems. 
If the planets started with a smaller semi-major axis
$a_1(0)=0.6\rm AU$, as shown in Figure\  \ref{a-e-distr}
then for each Jupiter in the radial range $0.1-0.5\rm AU$  our models produce
at least  ten more (mostly non-migrating) Jupiters within the range $0.5-1\rm AU$.
Contrary to this, observations yield roughly the same number of giant planets 
in these  two ranges of semi-major axes.

The lower panel of Figure\ \ref{a-e-distr} shows the final $a$ vs $e$ distribution of the simulated
systems  compared to the observed distribution.
This plot shows an additional important feature of the simulated  distribution which 
is difficult to reconcile 
 with observations: the lack of low-$e$ 
WJs at $a_1\lesssim 0.3\rm AU$. In fact,  none of our models produced 
a  WJ with $e\lesssim 0.3$ at these radii,
while $\sim 80\%$ of known planets with semi-major axis $0.25\leq
a_1\leq 0.1$ have eccentricities $\lesssim 0.3$. Evidently
giant planets observed  in this region of parameter space are
unlikely  to have formed through secular migration
induced by an outer perturber planet. 
This latter point is further discussed in the next section.

\begin{figure}
\centering
\includegraphics[width=2.9in,angle=270.]{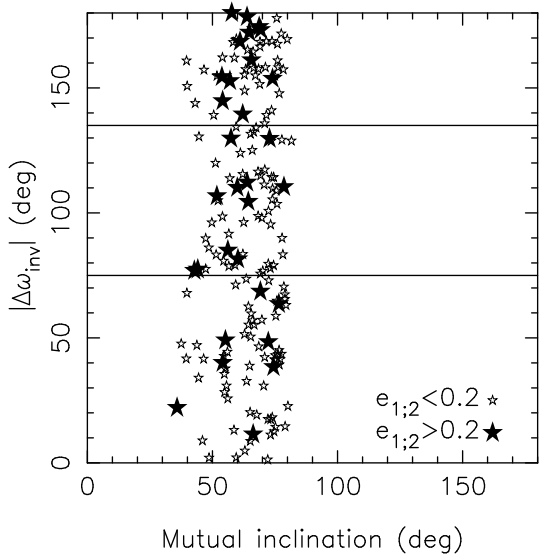}
\caption{Difference in apsidal longitudes , $\Delta \omega_{\rm inv}=\omega_1-\omega_2$,  as a function of mutual inclination 
for the WJs formed in our secular integrations. Filled-star symbols 
are the sub-set of systems comprising  an eccentric WJ with eccentric outer companion.
$e_1$ and $e_2$ are the final value of the eccentricity of inner and outer planets, i.e. those at the 
 time at which the apsidal longitudes were computed.
} \label{am-dis}
\end{figure}

\citet{2014Sci...346..212D} showed that eccentric WJs with
eccentric outer giant companions have apsidal separations which cluster near
$90^\circ$.  \citet{2014Sci...346..212D} interpreted this as signature of
mutual inclinations  being between $35-65^\circ$, 
favoring  LK migration as the mechanism for the formation of the inner planets.
 In Figure\ \ref{am-dis} we show the distribution of
the apsidal misalignment, $\Delta \omega_{\rm inv}=\omega_1-\omega_2$ , as a function of mutual inclination for
the migrating WJs in our secular integrations. 
Interestingly, our simulated planets do not show any significant  clustering 
around $\Delta \omega_{\rm inv}=90^\circ$. The WJs  formed in our models
have  mutual inclinations in the range $40-80^\circ$, but their
$\Delta \omega_{\rm inv}$ appear to be  uniformly distributed.
These results suggest
some other process, other than LK oscillations, as responsible 
 for the near orthogonality of apsides exhibited by the observed WJs.

\begin{figure*}
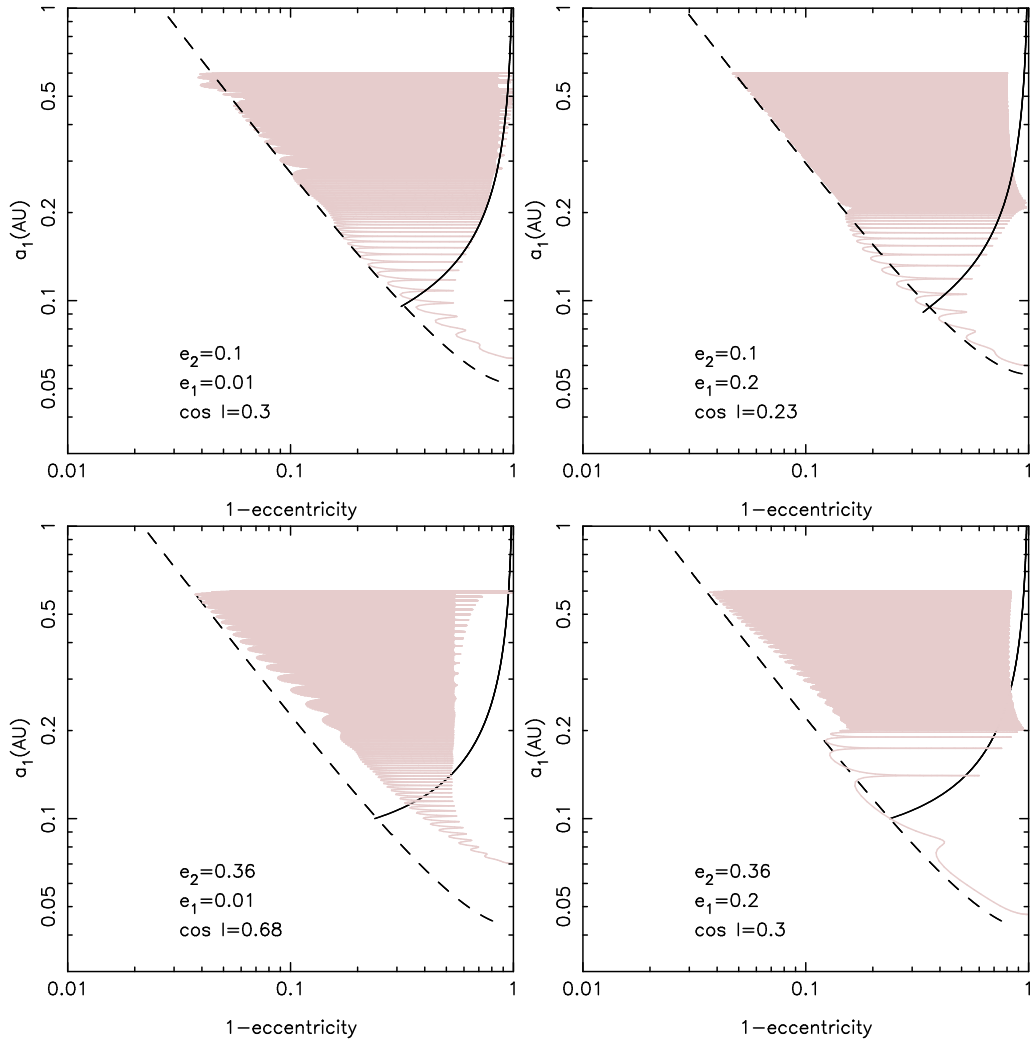

\centering
\includegraphics[width=2.7in,angle=270.]{Fig6a.eps}
\includegraphics[width=2.7in,angle=270.]{Fig6b.eps}\\
\includegraphics[width=2.7in,angle=270.]{Fig6c.eps}
\includegraphics[width=2.7in,angle=270.]{Fig6d.eps}
\caption{Orbital evolution of migrating planets obtained with the
  secular equations of  motion. The  solid lines 
give the minimum eccentricity attained during the LK oscillations 
as predicted by Equation\ (\ref{app-circ});
dashed lines are lines of constant   angular momentum which give the
value of $\ell_-$  we have used to compute the solid line.
In the lower panels the eccentricity of 
the outer planet is $e_2=0.36$, which causes the octupole order terms
to become important to the evolution.
In all panels the inner planet argument of periapsis was set to $\omega_1=\pi/2$
and the mass of the outer planet was $17M_{\rm Jupiter}$; $\omega_2$
was set equal to (in radians)  $3$ (upper left), $0.013$ (upper right and lower left) and
 $2.6$ (lower right). 
}\label{evo-oct}
\end{figure*}

\subsection{Evolution towards low semi-major axes}\label{dmcs}
As shown in Figure \ref{a-e-distr} a high-$e$ LK migration scenario 
for  WJs fails at producing 
systems with a low value of $a$ and $e$ that 
are indeed quite numerous among the observed systems. In fact none of our
simulations produced a system in the region where, for example, HD-38529b and
HD-13908b are observed ($0.1-0.3\rm AU$), disfavoring a LK induced migration scenario
for the formation of these planets.
As discussed below, the reason for this is the reduced range of eccentricity oscillations
 due to Schwarzschild precession as the planet semi-major axis shrinks
due to tidal dissipation. 

At the quadrupole order level and in the test particle limit 
the quantity
\begin{equation}\label{hamiltonian}
H=\ell^2+\sin^2I\left(\ell^2+5e_1^2\sin^2 \omega \right)-\frac{k}{\ell} \ ,
\end{equation}
with $\ell=\sqrt{1-e_1^2}$, is an integral of motion as it differs from the system Hamiltonian 
only by a constant \citep[e.g.,][]{david-book}. The 
third term in the right hand side of Equation\ (\ref{hamiltonian})
represents the extra Schwarzschild precession term where
\begin{equation}
k=8\frac{M_\star}{M_2}\frac{r_ga_2^3}{a_1^4}\left(1-e_2^2\right)^{3/2}
\ ,
\end{equation}
and $r_g=G M_\star/c^2$.

{
From the conservation of $H$ and $\ell_z= \ell\cos\ I$ we
can derive a relation between the maximum ($\ell_+$) 
and minimum ($\ell_-$) angular momentum attained during a LK 
oscillation for a given value of $\ell_z$. 
When  the argument of periapsis $\omega$ librates around $\pi/2$
the maximum and minimum
values of $\ell$ are related through the equation:
\begin{equation} \label{lib}
\ell_+^2\ell_-^2={\frac{5}{3}}\ell_z^2 +\frac{k}{3}\left(\frac{\ell_+-\ell_-}{\ell_+^2-\ell_-^2}\right)\ell_+\ell_-.
\end{equation}
As the planet semi-major axis decreases due to tidal dissipation,
relativistic precession will increase the portion of parameter space available for
circulation at expenses of libration,
gradually  pushing an initially librating orbit toward the separatrix at which $\ell_+=1$,
and finally  onto a  circulating orbit \citep[see also][for a similar analysis]{2002ApJ...578..775B,2015arXiv151008918A}.
 
 The distance at which 
the fixed point does not longer
exist 
 is found by setting $\omega=\pi/2$
and $\dot{\omega}=0$, which yields:
\begin{eqnarray}\label{fpp}
\tilde{a}&=&\left[4\frac{M_\star}{M_2}\frac{r_ga_2^3(1-e_2^2)^{3/2}}{3-5\ell_z^2}
\right]^{1/4}\approx 0.2\left(M_\star \over M_\odot\right)^{1/2} \\
&& \left( M_2\over M_{\rm Jupiter}\right)^{-1/4}
\left(a_2 \sqrt{1-e_2^2}  \over 3\rm AU\right)^{3/4}
\left(2{a_1\over a_1-a_{\rm cr}}\right) ^{1/4} \rm AU\ \nonumber  ,
\end{eqnarray}
where we have used the fact that $\ell_z\approx \sqrt{3/5}\ell_-$
(from $\ell_z=\cos (I)
\ell$ and $\cos (I)\approx \sqrt{3/5}$ at $\ell_-$) 
 and
set $\ell_-=a_{\rm cr}/a_1$ as required for efficient tidal dissipation to occur.
Below $\tilde{a}$ librating solutions do not longer exist.

After $\omega$ starts circulating $\omega=\pi/2$ at $\ell=\ell_-$
and $\omega=0$ at $\ell=\ell_+$, which leads to the relation
\begin{equation}\label{circ}
\ell_+^2=
\frac{5}{2}\left(1+\ell_z^2-\frac{3}{5}\ell_-^2-\frac{\ell_z^2}{\ell_-^2}
\right)
+\frac{k}{2}\left(\frac{1}{\ell_+}-\frac{1}{\ell_-}\right) \ .
\end{equation}
According to 
Equation\  (\ref{circ}) and for  $a_1<\tilde{a}$,
$\ell_+$ must become smaller as $a_1$ decreases,  thereby pushing the planet away from
the region of small $a_1$ and $e_1$.
}

A good approximation to Equation\ (\ref{circ}) can be obtained 
by  noting that $\ell_z\approx \sqrt{3/5}\ell_- $ 
and $k/\ell_+\approx k$ (from $\ell_+\approx 1$), which leads to the simpler relation
\begin{equation}\label{app-circ}
\ell_+^2\approx 1+\frac{k}{2}\left(1-\frac{1}{\ell_-}\right) \ ,
\end{equation}
for circulating orbits.
Although quite simplified and reasonable only for an orbit close to
the separatrix, Equation\ (\ref{app-circ}) was found to reproduce 
the results of numerical simulations fairly well.
A few  example systems are shown in Figure \ref{evo-oct}.
The dashed curve in the figures that demarcate the $\ell_-$
envelope is a curve of constant angular momentum:
$\sqrt{a_1}\ell$. $\ell_-$ tracks this curves because tidal
dissipation occurs mostly at $\ell=\ell_-$.
In the upper left panel the inner planet argument of periapsis  is
initially circulating and the $\ell_+$ value steadily  decreases  with time.
In the upper 
right panel instead $\omega$  is initially librating.
From Equation (\ref{lib}) we see that  the inclusion of the extra Schwarzschild precession term
will tend to increase  $\ell_+$ as the orbit shrinks. Accordingly, from 
Figure \ref{evo-oct} we see that as the orbit decays  $\ell_+$ increases 
until it crosses the
separatrix  at $a_1\approx 0.2\rm AU$
where  $\ell_+\approx 1$. Then the  $\ell_+$ envelope is set by  the separatrix, as modified by
Schwarzschild precession and it is approximately equal to the value 
given by Equation\ (\ref{app-circ}) after $\omega$ starts circulating.

\subsubsection{Effect of octupole order terms}
If the orbit of the outer planet has a substantial eccentricity
(typically $\gtrsim 0.1$),
then the octupole order terms can cause the evolution of the inner
planet orbit to deviate significantly from the simple model depicted above.
However, the distribution shown in Figure \ref{a-e-distr}, which was
obtained with the octupole secular code,  suggests
 that even when  higher order terms are included the innermost planet orbit 
 keeps away from the region $a_1\lesssim 0.3\rm AU$, $e_1\lesssim
 0.3$, for $a_2\geq 0.6\rm AU$.

The octupole order terms have two main effects:
(i) the high eccentricity part of the envelope
can deviate significantly from  a line of constant angular momentum
(e.g., lower panels of Figure\ \ref{evo-oct});
(ii) if the inner planet argument of periapsis is initially librating, as the
semi-major axis decreases due to tidal friction  the orbit eventually 
crosses  the LK separatrix where dynamical chaos can drive the orbital eccentricity to very
high values. 
When this happens the planet orbit will tend to
``freeze'' at  higher values of $e$ so that a HJ will
promptly form (see the bottom right panel of Figure\ \ref{evo-oct}).

\subsection{Multi-planet systems and secular chaos}\label{sc}
As mentioned before, another method to excite eccentricities is via
secular chaos in multi-planet systems \citep{2011ApJ...735..109W}.
Here we show that the conclusions drawn in this section likely 
applies to such type of systems as well.

In systems that host more than two giant planets, the planets need not be
close companions, and need not be initially highly eccentric and/or
inclined to excite the innermost planet eccentricity  to high values, and
potentially produce a WJ \citep[e.g.,][]{2015MNRAS.449.4221H}. 
However, population synthesis studies with plausible assumptions
(Hamers et al. in prep., Antonini et al. in prep.)  find that almost
no WJs are produced,  whereas HJs are produced in more
significant numbers (up to a few per cent).

\begin{figure}
\centering
\includegraphics[width=3.3in,angle=0.]{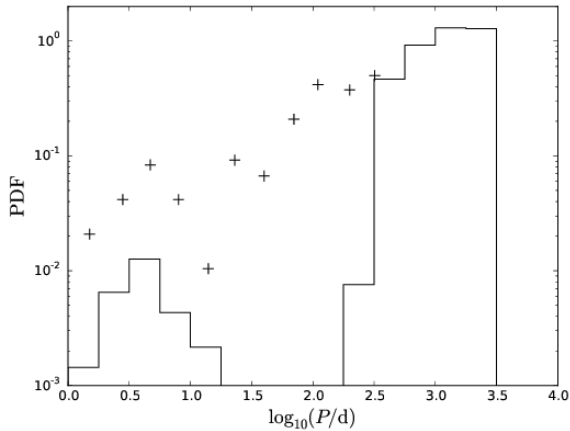}
\caption{Period distributions from secular calculations of high-$e$
  migrations in multi-planet systems, 
with 3 to 5 planets (Hamers et al. in prep). Crossed symbols show the
observational data from \citet{2015arXiv151100643S}, normalized to match the
simulated  distribution at $\mathrm{log}_{10}(P/\mathrm{d}) \approx 2.6$ .}\label{sec-c}
\end{figure}

In Figure\ \ref{sec-c} we show  the period distribution of a large set of
multi-planet system integrations based on the orbit-averaged code
described in \citet{2015arXiv151100944H}. 
In these secular simulations, the number of planets was chosen between 3
and 5, and the 
semi-major axes were sampled linearly between 1-4, 6-10, 15-30, 45-50
and 60-100 AU for the 3 to 5 planets, respectively.
 The stellar mass was set to $1\, \mathrm{M}_\odot$, and the planetary
 masses were sampled
randomly  between 0.5 and 5 $M_\mathrm{Jupiter}$. 
 The arguments of periapsis and longitudes of the ascending nodes
 were sampled randomly. The apsidal motion
constant was set to 0.28. 
The time-lag constant was set to $\tau=66\rm sec$,
and the innermost 
planet radius was  to either 1 or 1.5 $R_\mathrm{Jupiter}$; 
The inclinations and
 eccentricities (in units of radians) were
 sampled from a Rayleigh distribution with an rms width of either 10
 or 15 degrees ($\approx 0.18$ or $\approx 0.35$ radians).

In total we integrated $10000$ systems up to a maximum integration time
of $10\ $Gyr.
In our analysis we rejected all
systems in which the planet orbits crossed during the integration or in which
the inner planet collided with the star.
We also tried different values of $\tau(=0.66,\ {\rm and}\ 0.066\rm sec)$ but
found this not to affect our conclusion:
similar to the results of the two-planet system integrations described
above, and  in stark contrast with the observations, 
a small number of WJs is produced compared to HJs. 
Indeed our multi-planet simulations produce essentially  no Jupiter in the
period valley as can be seen in Figure\ \ref{sec-c}.
We conclude that these models as well, greatly underpredict the number of giant planets 
observed in the period valley.

\begin{figure}
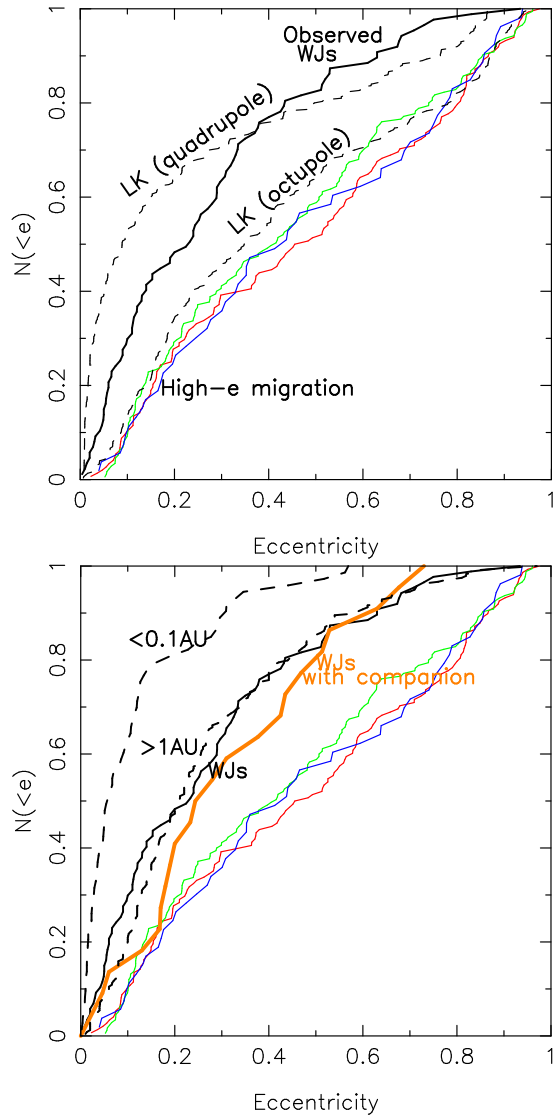

\centering
\includegraphics[width=2.9in,angle=270.]{Fig8a.eps}\\
\includegraphics[width=2.9in,angle=270.]{Fig8b.eps}
\caption{Cumulative eccentricity distribution of 
migrating planets  from our two-planets simulations
(green, red and blue lines) compared to the observed eccentricity distribution of WJs
(black solid lines).
All the high-$e$ migration models
produce an eccentricity distribution which is significantly different
from the observed distribution.
In the top panel the dashed curves give the eccentricity distribution
of a planet undergoing LK oscillations computed using the secular
equations of motion and at the quadrupole (upper curve) and octupole
(lower curve)  level of approximation.
In the bottom panels the black dashed curves are the 
eccentricity distributions of observed Jupiters  in the indicated range
of semi-major axes. 
Orange solid line is for the sub-sample of WJs with one outer companion.
The model predictions do not take into account  eccentricity-dependent
selection effects which might become important at $e\gtrsim 0.8$.
} \label{e-dis}
\end{figure}

\section{Eccentricity distribution}\label{ecc-d}
 Figure \ref{e-dis} compares the eccentricity  distribution of our
 simulated systems to the distribution of observed WJs. 
The simulated models yield an eccentricity distribution for the
migrating planets which is nearly uniform ($N(<e)\sim e$). 
A comparison of these two populations show that 
our migration models produce too many 
highly eccentric WJs to be consistent with observations.
The oscillations required to produce inward migration 
result in more  eccentric planets  than
observed so that the simulated distribution for the migrating population is inconsistent with
the observed eccentricity distribution \citep[see also][]{2015ApJ...798...66D,2016MNRAS.455.1538F}.
The discrepancy of the migration model with observations 
is therefore due to the significant fraction  of migrating 
WJs with high eccentricity, while only a few observed WJs are on  high
eccentricity orbits (see Figure \ref{a-e-distr}).
In fact, as also noted above, a large number of WJs 
have an eccentricity that is close to zero.

In  Figure  \ref{e-dis}  we show the model eccentricity distributions
starting from different  initial conditions (models A3, B3 and C3-c in
Table 2).
These models all produce a similar final eccentricity distribution
demonstrating that our conclusions are quite robust and
do not depend significantly on the choice we made for the initial
conditions. Using the  Kolmogorov-Smirnov (K-S) test gives $p$-values 
in the range  $10^{-3}-10^{-4}$, indicating
that the synthetic  and observed $e$-distributions  are unlikely to be drawn from the same 
 population. 

The dashed curves  in the top panel  of Figure \ref{a-e-distr}
give the eccentricity distribution of a planet
undergoing  LK oscillations that we
computed using the secular equations of motion at the quadrupole 
(upper curve) and octupole (lower curve) level of approximation. 
The planet was
placed at $a_1(0)=0.5\rm AU$ with a negligible initial eccentricity ($e_1(0)=
0.01$). The outer planet had a mass of $5M_{\rm Jupiter}$ and was
placed at $a_2=6\rm AU$ with $e_2(0)= 0.2$. The mutual inclination was
set to $70^\circ$.  At the quadrupole level of approximation we can
simply  derive the eccentricity distribution   as
$dN/de = dt/de\sim 1/e_1$ so that $N\sim \ln e$ -- this follows from the fact 
that in a mixed ensemble the number of planets $\Delta N$ in the interval
$e \sim e + \Delta e$ is proportional to  $\Delta t$.
As expected,  this form matches quite well the  eccentricity distribution given by the 
upper dashed  curve in Figure \ref{a-e-distr}, but does not provide a good
mach to either  the observed eccentricity distribution or to the model distribution.
The distribution of the
migrating Jupiters in our two-planet simulations is  instead similar to that of a
population of planets undergoing LK oscillations with a
non-negligible contribution from the octupole potential (lower dashed curve).
We conclude that the dynamical evolution of the planets in our simulations is significantly affected by the 
octupole order terms. The main effect of the octupole
order terms is  to  skew the eccentricity distribution of WJs towards
higher values.

In the lower panel of Figure \ref{e-dis} we compare the eccentricity
distribution of planets within the range of semi-major axes:
$0.1-1\rm AU$, $\leq0.1\rm AU$ and $\ge 1\rm AU$.
Interestingly, we find that the eccentricity distribution
of giant planets  in the radial range $0.1-1\rm AU$ is consistent with
that of planets at radii larger than $1\rm AU$.
A K-S test on these distributions gave a $p$-value of 0.31 implying
that the two samples are consistent with being taken from the same
distribution. For comparison, the same test between the distribution of 
planets with $a>1\rm AU$  and giant  planets at $a<0.1\rm AU$
gave  a $p$-value of $\sim 10^{-5}$.
Hence the observed $e$-distribution provides no
evidence for differences in the 
eccentricity distribution of WJs and Jupiters outside $1\rm AU$ that
is expected on the basis of theoretical models.
In addition, comparing the eccentricity
distribution of all WJs with that of only WJs with one outer companion
(orange curve) shows that these two distributions are not significantly
different from each other.
All these results point either to disk migration \citep{1980ApJ...241..425G}  or to 
in-situ formation \citep{2015arXiv151109157B,2016ApJ...817L..17B} 
for the origin of WJs rather than secular migration processes such as the
LK mechanism \citep{2014Sci...346..212D} or secular chaos \citep{2011ApJ...735..109W}.

\section{Discussions and Conclusions}
We have considered the population of giant planets in the period valley. These
planets, often referred to as WJs, have orbital periods larger than $10\rm days$ but are interior 
to the peak of giant planet frequency observed at $\approx 1\rm AU$
\citep{2015arXiv151100643S}.
It has been argued that such planets might not be able to form in-situ.
In a widely discussed model for the
formation of these planets, large amplitude eccentricity
oscillations induced by an external perturber are 
followed by efficient tidal dissipation which causes
the orbit of the inner planet to shrink during close passages by the
host star
\citep[e.g.,][]{2014ApJ...781L...5D,2014Sci...346..212D,2016MNRAS.455.1538F}.

Before summarizing our results we briefly  address the importance of selection effects. 
In fact, when comparing the predictions of our models to observations we have
so far neglected the fact that observations might be biased 
against for example  orbits with  high
eccentricity and/or large semi-major axis. This might affect the inference of the
intrinsic orbital distribution of the observed planets
\citep[e.g.,][]{2012ApJ...750..106S}.

We have considered mostly  RV data from  the exoplanets.org database.
RV data might be  biased against the detection of longer period
planets, although this effect is likely to be small for the 
semi-major axis range  considered here $a<1\rm AU$ \citep{2004MNRAS.354.1165C}.
More important might be the bias against the
detection of eccentric planets which would affect the  distributions
of Figure \ref{e-dis}.
In fact, the sparse sampling of an orbit with a high-eccentricity
can miss the reflex velocity signal near periapsis, leading to
non-detection of planets that would be detected at the same semi-major
axis and lower eccentricity \citep{2004MNRAS.354.1165C}.
We believe however that these effects should be relatively small 
 since our sample is restricted to giant planets with relatively large
 masses  ($M\sin I>0.5 M_{\rm Jupiter}$)  and small semi-major axes
 ($<1\rm AU$),  which should be relatively easy to detect.
We note also that other studies which have taken into account  such
selection effects have reported 
results similar  to ours,  pointing out the excess of highly eccentric WJs predicted by
 high-$e$ scenarios compared to the observed  
distribution \citep{2015ApJ...798...66D,2016MNRAS.455.1538F}.

We also note that we have limited the parameter space of our simulations by keeping the perturber 
mass within $17\rm M_{\rm Jupiter}$. However,
the WJs for which there is no observational evidence for
a Jupiter companion might be migrating due to interactions with a 
distant stellar companion. \citet{2015ApJ...799...27P} and \citet{2015arXiv151008918A} conducted octupole-level
population synthesis studies of giant planets migrating through the LK mechanism due
to a stellar companion and friction due to tides.
Although their initial conditions are  different from ours,
the fraction of migrating planets obtained in these studies and
their orbital distribution are comparable to what is obtained
in our study. For example, the fraction ($\sim 1\%$) and orbits of migrating
planets displayed in Figure 10 of \citet{2015ApJ...799...27P} are
clearly similar to those shown in our Figure\ \ref{a-e-distr}. 
This suggests that while the perturber plays the
fundamental  role in inducing the planetary LK oscillations, 
the perturber properties are likely to not impact our general results.
The main results of our paper should therefore  apply also to the case in which the 
LK oscillations are induced by a stellar companion rather than an
outer Jovian companion.

In conclusion we have presented a numerical study
of the dynamics of giant planets with close friends.
We used both a secular code based on orbit average equations of motion 
as well as direct three body integrations to address whether 
the giant planets observed in the semi-major axis range $0.1-1\rm AU$
could have been formed farther out and then migrated to these radii
through secular migration processes such as LK cycles or secular chaos.  
The main results of our study are summarized below:
\begin{itemize}
\item[1] according to  the high-$e$ migration hypothesis, HJs and WJs
formed originally at $\gtrsim 1\rm AU$ distance from their host star.
In order to test this hypothesis 
we addressed whether the observed Jupiter pairs 
hosting  HJs and  WJs would be dynamically stable if the inner planet
was placed on an eccentric orbit (as required for efficient tidal dissipation) at
$\gtrsim 1\rm AU$. According to stability criteria that we have taken from the literature,
only four of the twenty-four observed systems
would be dynamically stable at these radii with fourteen of them being on
intersecting orbits. As we also confirmed by direct 
integrations, if a planet pairs is unstable, it does not 
lead to the formation of tidally migrating planets but rather to
collisions with the host star or planet ejections. These findings
point against a high-$e$ migration
scenario from $\gtrsim 1\rm AU$  for the formation of most  observed systems.
  
\item[2] 
We showed that  high-$e$ migration models for WJs produce 
a period distribution that is not consistent with observations.
By comparing the ratio of WJs to CJs we found in our
 simulations to that ratio as observed, we infer that $\lesssim 10\%$ of all gas giants
observed at $0.1-1\rm AU$ from their stellar host could have formed
through high-$e$ migration LK processes. 
Preliminary simulations of systems containing three to five
planets suggest that the
fraction of WJs produced in multi-planet systems is likely to be small
as well.

\item[3] Our analysis shows that high-$e$ migration processes
tend to  produce more HJs than WJs. Accordingly,
 for any detected WJ with a close companion 
there should be at least an equal amount of detected HJs also with a
close companion. 
In contrast with this prediction, 
and restricting ourself to Jupiters discovered through  RV  surveys, 
we find that only  2 HJs  have a detected outer giant
companion within $\lesssim 5\rm AU$, while
20 of the 74 period valley gas giants do. This  points towards a different 
formation history for the two populations of planets.

\item[4] 
Using both numerical and analytic techniques we have shown that 
a tidal migration model produces an eccentricity distribution for the
migrating planets that also appears to be  inconsistent with observations.
The oscillations required to produce inward
migration  tend to excite the eccentricities of migrating planets to values higher 
than those observed.

\item[5]We showed that the observed eccentricity distribution
of giant planets  in the radial range $0.1-1\rm AU$ is consistent with
the eccentricity distribution for planets at radii larger than $1\rm
AU$. This might indicate a close relation between the two populations of planets 
and perhaps a common formation history.

\end{itemize}

Based on these results, we conclude that rather than starting on highly-eccentric orbits 
with orbital periods above 1 year, the Jupiters observed in the radial
range  $0.1-1\rm AU$ from their stellar host are likely to have
reached the region where  they are observed today without tidal
circularization. Tidal migration following  planet-planet scattering is also disfavored given
that we would not expect it to typically result in close and mildly
eccentric companions to WJs \citep{2008ApJ...678..498N,2012ApJ...751..119B}.

Where do the WJs come from then?
Our results may indicate that disk migration is the
dominant channel for  producing WJs. Alternatively, they might have formed
in-situ \citep{2016arXiv160105095H}, i.e., they underwent runaway gas accretion, from 
originally  low-mass closely packed planets \citep[e.g.,][]{2014ApJ...797...95L}.
 However, both
disk migration and in-situ formation are
more likely to produce planets on nearly circular and low inclined orbits, so it
remains to be explained how in these scenarios the planets’ eccentricities could 
have been excited to the observed values. 
A possibility is that the eccentricities were excited through secular
chaos \citep{2011ApJ...735..109W}, 
which could imply the presence of one or multiple 
still undetected planet companions to the numerous planet pairs observed in the period valley. \footnote{As this paper was submitted, 
we became aware of a similar study by \citet{2016arXiv160400010P}.
Based on a population-synthesis study similar to ours 
\citet{2016arXiv160400010P}
conclude that 
 high-eccentricity migration excited by an outer planetary companion  can account 
at most for $\sim20\%$ of the warm Jupiters, a fraction 
twice as large as the upper limit found here.}

\bigskip
We thank Bekki Dawson and Eugene Chiang for useful 
suggestions, and the referee for their detailed  comments 
that helped to improve the paper.
FA acknowledges support from 
a CIERA postdoctoral fellowship at Northwestern
University. ASH was supported by the Netherlands Research
Council NWO (grants 639.073.803 [VICI], 614.061.608 [AMUSE] and
612.071.305 [LGM]) 
and the Netherlands Research School for Astronomy (NOVA).
YL acknowledges grant  AST-1352369 from NSF and NNX14AD21G from NASA.

  \end{document}